\RequirePackage{tikz} 

\documentclass[pdflatex, sn-apa]{sn-jnl}

\usepackage{cmap}						
\usepackage[T1]{fontenc}				
\usepackage[nameinlink]{cleveref}
\usepackage{csquotes}  
\usepackage{pdflscape} 

\newcommand{\numcommandhistories}{8834}
\newcommand{\numtrainees}{113}
\newcommand{\numsessions}{18}
\newcommand{\kypo}{KYPO Cyber Range Platform}
\newcommand{\creator}{Cyber Sandbox Creator}

\jyear{2022}
\newcommand\copyrighttext{%
  \footnotesize This version of the article has been accepted for publication, after peer review, but is not the Version of Record and does not reflect post-acceptance improvements or any corrections. The Version of Record is available online at: \url{https://doi.org/10.1007/s10639-022-10954-4}. Use of this Accepted Version is subject to the publisher’s Accepted Manuscript terms of use: \url{https://www.springernature.com/gp/open-research/policies/accepted-manuscript-terms}
  \smallskip
  
  \textcopyright\ 2022. Please cite this article as follows: V. Švábenský, J. Vykopal, P. Čeleda, K. Tkáčik, and D. Popovič: \textit{Student Assessment in Cybersecurity Training Automated by Pattern Mining and Clustering}, Springer Education and Information Technologies, 2022, ISSN 1360-2357, DOI: \url{https://doi.org/10.1007/s10639-022-10954-4}}
\newcommand\copyrightnotice{%
\begin{tikzpicture}[remember picture,overlay]
\node[anchor=north,yshift=-24pt] at (current page.north) {\fbox{\parbox{\dimexpr\textwidth-\fboxsep-\fboxrule\relax}{\copyrighttext}}};
\end{tikzpicture}%
}

\begin{document}

\title[\footnotesize Student Assessment in Cybersecurity Training Automated by Pattern Mining and Clustering]{Student Assessment in Cybersecurity Training Automated by Pattern Mining and~Clustering

\copyrightnotice\vspace*{-0.75cm}}

\author*[1,2]{\fnm{Valdemar} \sur{Švábenský}}\email{svabensky@ics.muni.cz}
\author[1]{\fnm{Jan} \sur{Vykopal}}\email{vykopal@ics.muni.cz}
\author[1]{\fnm{Pavel} \sur{Čeleda}}\email{celeda@ics.muni.cz}
\author[2]{\fnm{Kristián} \sur{Tkáčik}}\email{tkacikk@mail.muni.cz}
\author[2]{\fnm{Daniel} \sur{Popovič}}\email{popovic@mail.muni.cz}

\affil[1]{\orgdiv{Institute of Computer Science}, \orgname{Masaryk University}, \orgaddress{\street{Šumavská~15}, \city{Brno}, \postcode{60200}, \country{Czech Republic}}}
\affil[2]{\orgdiv{Faculty of Informatics}, \orgname{Masaryk University}, \orgaddress{\street{Botanická 68a}, \city{Brno}, \postcode{60200}, \country{Czech Republic}}}

\abstract{Hands-on cybersecurity training allows students and professionals to practice various tools and improve their technical skills. The training occurs in an interactive learning environment that enables completing sophisticated tasks in full-fledged operating systems, networks, and applications. During the training, the learning environment allows collecting data about trainees' interactions with the environment, such as their usage of command-line tools. These data contain patterns indicative of trainees' learning processes, and revealing them allows to assess the trainees and provide feedback to help them learn. However, automated analysis of these data is challenging. The training tasks feature complex problem-solving, and many different solution approaches are possible. Moreover, the trainees generate vast amounts of interaction data. This paper explores a dataset from \numsessions\ cybersecurity training sessions using data mining and machine learning techniques. We employed pattern mining and clustering to analyze \numcommandhistories\ commands collected from \numtrainees\ trainees, revealing their typical behavior, mistakes, solution strategies, and difficult training stages. Pattern mining proved suitable in capturing timing information and tool usage frequency. Clustering underlined that many trainees often face the same issues, which can be addressed by targeted scaffolding. Our results show that data mining methods are suitable for analyzing cybersecurity training data. Educational researchers and practitioners can apply these methods in their contexts to assess trainees, support them, and improve the training design. Artifacts associated with this research are publicly available.}


\maketitle

\vspace*{-0.25cm}

\section{Introduction}
\label{sec:intro}

Cybersecurity professionals are needed across the globe to counter the ubiquitous cyber threats. To meet the increasing demand for cybersecurity experts~\citep{isc2}, effective training is essential. Such training must include hands-on components and provide practical experience in authentic settings. This includes using a variety of tools for cybersecurity operations, such as host configuration, hardening, and penetration testing.

Cybersecurity experts use tools with graphical user interfaces as well as command-line tools. Within the scope of our research, we focus on the latter, since command-line tools represent an important component of cybersecurity practice. Cyber attackers use them to perform sophisticated attacks, which cyber defenders need to understand to mitigate advanced threats. In addition, various command-line tools are used to configure computer systems securely.

\subsection{Research Problem Statement}

Our research focuses on \textit{supporting automated assessment in the context of hands-on cybersecurity training}. Here, we explain the motivation for our research, illustrate the problem with a simple example, justify why the problem is hard to address, and summarize the gaps in the current literature.

\subsubsection*{Why Is Student Assessment Necessary?}

Educational assessment is a crucial aspect of training~\citep{handbook-CER14}. It enables teachers (\textit{instructors}) to better understand the actions of their students (\textit{trainees}). Specifically, in-depth assessment shows what each student did well, what could be improved, and whether the student progressed through the training as expected.

Based on insights from the assessment, teachers can adapt their class, provide students with feedback to support their learning, or evaluate their level of knowledge. The assessment also shows potential issues in the training design, enabling to fix them and further improve the effectiveness of the training.

\subsubsection*{How Can Students Be Assessed?}

Like most applied computing skills, cybersecurity is usually practiced hands-on in computer-supported interactive learning environments. These are physical or virtual platforms that provide computer hosts with full-fledged operating systems, networks, and applications for training.

Advanced interactive learning environments allow collecting data about the students' actions, such as their usage of command-line tools. These student interaction data authentically capture learning processes. Therefore, they can be transformed into educational insights and exploited for assessment.

As an example, consider the two command histories from cybersecurity training shown in \Cref{fig:1b-bruteforce,fig:1c-help}. They belong to two students who attempted to crack a password to a ZIP archive using the \texttt{fcrackzip} utility in Linux. Each command is prefixed by the timestamp of its execution. Based on the analysis of these student data, the instructor can see that each student needs help with a specific and different aspect of the training.

\begin{figure}[!ht]
\begin{center}
\small
\begin{verbatim}
    12:27:41 fcrackzip -b -u file.zip
    12:28:25 fcrackzip -b -D -u file.zip
    (repeated 5 times)
    12:29:17 fcrackzip -b -u /file.zip
    12:29:24 fcrackzip -b -u
    12:29:34 fcrackzip -b -u /file.zip
    12:30:13 fcrackzip -D -u -v file.zip
    12:30:51 fcrackzip -D -u -v file.zip
    12:30:55 fcrackzip -b -u -v file.zip
    12:31:18 fcrackzip -c -u -v file.zip
    (9 more executions with different argument combinations)
    12:33:02 fcrackzip -b -u -v file.zip
    12:37:22 fcrackzip -v -u -D fasttrack.txt file.zip
    
\end{verbatim}
\end{center}
\caption{The first student ran the cracking tool 24 times within an approximately 5-minute time frame, with various combinations of arguments, often repeating the previous (incorrect) combinations. After that, the student stopped for 4 minutes, probably to find help, and executed a correct command.}
\label{fig:1b-bruteforce}
\end{figure}

\begin{figure}[!ht]
\begin{center}
\small
\begin{verbatim}
    20:48:49 fcrackzip.exe-help
    20:48:55 fcrackzip.exe -help
    20:49:02 fcrackzip.exe --help
    20:49:04 fcrackzip
    20:49:08 --help
    20:49:13 fcrackzip --help
    20:50:44 sudo apt-get install fcrackzip
    20:52:51 fcrackzip -u -D -p fasttrack.txt file.zip
    
\end{verbatim}
\end{center}
\caption{The second student assumed that the tool had the \texttt{.exe} suffix of Windows OS executables, which does not apply to Linux OS. The student was apparently unfamiliar with Linux or the cracking tool but then instantly executed a correct command without any previous incorrect tries. We can assume that they received outside help.}
\label{fig:1c-help}
\end{figure}

\subsubsection*{Why Is Assessment Difficult?}

In-depth assessment of cybersecurity training is difficult for four main reasons.
\begin{enumerate}
    \item \textit{The training is complex.} The tasks require high-order problem solving and may have many different correct solutions. Therefore, the assessment is much more complex than assessing simple tasks such as memorizing facts.
    \item \textit{Each student is unique.} Every student has different previous knowledge, experience, motivation, and approach to learning. As a result, students adopt different strategies to solve the tasks. This is natural, but it further complicates the conditions for automatically assessing hands-on tasks.
    \item \textit{Students generate a lot of data.} During the training, even a class that is relatively small (10--20 students) and time-constrained (1--2 hours) can generate hundreds of data records. As a result, manually processing these data becomes quickly infeasible.
    \item \textit{The assessment process is not straightforward.} It is unclear how to transform the raw data from training into educational insights useful for assessment. As the examples in \Cref{fig:1b-bruteforce,fig:1c-help} demonstrated, even a relatively constrained assignment can generate various data for assessment.
\end{enumerate}

\subsubsection*{The Need for Research}

Traditionally, educational researchers and practitioners assessed student data manually. However, due to the difficulties described above, a manual transformation of hands-on training data into educational insights is not viable~\citep{romero2020, fournier2017blog}. It is highly time-consuming, ineffective, and error-prone.

Automated assessment is more scalable and accurate. Therefore, it can be fruitful to leverage automated techniques, such as machine learning and data mining, for analyzing data from hands-on training~\citep{Palmer2019}. These techniques should transform the data from their raw form to an understandable representation, such as an overview of highlights or a visualization.

However, the review of current literature (see \Cref{sec:related-work} for details) identified several gaps in state of the art in this area:
\begin{itemize}
    \item As \cite{weiss2016} argued, current automated assessment is often superficial, judging only the (in)correctness of the solution. Only a few papers, such as by \cite{mirkovic2020}, have explored an in-depth assessment of student learning.
    \item To the best of our knowledge, no published research attempted to compare and evaluate the applicability of two different data mining methods on cybersecurity training data. Student assessment in cybersecurity has been explored from other perspectives, such as using numerical scoring metrics (see \cite{Maennel2017} for an example).
    \item Data mining algorithms have been used for assessment in other domains, such as programming~\citep{gao2021}, but it is unclear how to generalize these previous results to the cybersecurity context.
\end{itemize}

\subsection{Goals of This Research Paper}
\label{subsec:goal}

We seek to support automated assessment of students in hands-on training. In order to address the gaps in the literature, the assessment must satisfy the following criteria:
\begin{itemize}
    \item enable an in-depth understanding of students' actions,
    \item use methods that have not been researched in this context previously, and
    \item be evaluated on an authentic dataset from realistic training sessions. 
\end{itemize}

The domain of data mining offers many methods for the automated extraction of insights from raw data~\citep{fournier2017blog}. Two methods that satisfy the criteria above and will be explored in this paper are \textit{pattern mining} and \textit{clustering}. Pattern mining techniques, such as association rule mining and sequential pattern mining, can reveal interesting relationships in datasets~\citep{fournier2013blog}. Clustering, on the other hand, forms groups of data based on their similar characteristics~\citep{handbook-edm2010}. Evaluating these two techniques represents an original contribution to cybersecurity education and beyond.

\subsubsection*{Research Questions}

Our research is framed by two research questions related to student assessment in cybersecurity: \textit{What insights can we gather from command histories using pattern mining} (RQ1) \textit{and clustering} (RQ2)\textit{?} By \textit{insights}, we mean the following educational findings to support assessment:
\begin{itemize}
    \item trainees' approaches and strategies to solving the training tasks,
    \item common mistakes, misconceptions, and tools problematic for trainees,
    \item distinct types of trainees based on their actions and behavior, and
    \item issues in the training design and execution.
\end{itemize}

\subsubsection*{Expected Contributions of This Research}

Answering the research questions will be valuable for various stakeholders.
\begin{itemize}
    \item \textit{Cybersecurity instructors} can use the researched methods in their classes to gain new insights for assessing their students. Specific assessment use cases are detailed in \Cref{subsec:results-cmp} and \Cref{subsec:results-implications}.
    \item \textit{Researchers} can build upon this work by evaluating other data mining methods on similar datasets. This will contribute to the body of knowledge on assessment in cybersecurity training.
    \item \textit{Developers of cybersecurity training platforms} can integrate the researched methods of data collection and analysis into the interactive learning environments. This will support the goals of instructors and researchers.
\end{itemize}

Educational stakeholders from outside the cybersecurity domain can benefit from this research as well. Students of related computing disciplines, such as networking and operating systems administration, can generate similar data for assessment in hands-on classes. For students of other disciplines, the researched methods can be extended to process different data, such as clickstreams.

\subsection{How to Read This Paper}

Above, we defined three target groups who may be interested in this paper. Although we aim to address readers from a broad audience, we acknowledge that some sections of the paper are not relevant for everyone.
\Cref{sec:background} provides a brief background and therefore aims at \textit{researchers} who seek to understand the theory of the used methods. Other readers who are satisfied with a more high-level understanding may skip it.
\Cref{sec:related-work} reviews related studies, which is relevant for \textit{researchers} and \textit{instructors} interested in how the previous research results were applied to support teaching practice.
\Cref{sec:methods} details the used methods for the data collection and analysis. It is aimed mainly at \textit{researchers} and \textit{developers}, since it also includes technical details about the training platforms and data collection.
\Cref{sec:results} presents the findings and answers the research questions.
Finally, \Cref{sec:conclusion} concludes, summarizes our contributions, and proposes future work. These two sections are suitable for all readers.

\section{Background and Terminology of Data Mining}
\label{sec:background}

This section defines the key terms to familiarize the readers with basic data mining concepts. \textit{Data mining} is a field of computing that deals with extracting knowledge from data. Its purpose is to enable understanding of the data, gather new insights from them, and support decision-making based on this understanding~\citep{fournier2017survey, dm-concepts2011}. Out of the many data mining methods, we will focus on two of them: pattern mining (\Cref{sec:background:pattern}) and clustering (\Cref{sec:background:clustering}).

\subsection{Educational Data Mining and Learning Analytics}

\textit{Educational data mining} (EDM)~\citep{handbook-edm2010} and \textit{Learning analytics} (LA)~\citep{handbook-la2017} are two inter-related research areas that aim to understand and improve teaching and learning. The research in these areas focuses, for example, on student behavior, learning processes, assessment, and interactive learning environments. To achieve their aims, EDM/LA researchers collect and analyze data from educational settings.

\subsection{Pattern Mining}
\label{sec:background:pattern}

\textit{Pattern mining} automatically extracts previously hidden patterns in data. Its objective is to discover patterns that are easily interpretable by humans. We concentrate on two well-established pattern mining techniques: association rule mining (ARM) and sequential pattern mining (SPM)~\citep{fournier2013blog, fournier2017survey}.

\subsubsection*{Association Rule Mining}
\label{sec:background:arm}

\textit{Association rules} are patterns with the form of an \textit{if-then} statement. A rule $X \rightarrow Y$ says that if an \textit{item} $X$ occurs in a \textit{transaction} (a set of items), then so does $Y$~\citep{fournier2017survey, dm-concepts2011, handbook-edm2010}. In our case, an item may be a command submitted by a student, and a transaction may be a whole set of commands of that student. An association rule mined from a set of students' transactions may indicate that if a student used a command $X$, then they also used a command $Y$.

For each association rule $X \rightarrow Y$, we are typically interested in two metrics: its \textit{support} (relative occurrence among all the examined transactions) and \textit{confidence} (relative occurrence among the transactions that contain $X$).

Algorithms for mining association rules consider only rules that satisfy the user-defined thresholds for the minimal support and confidence, \texttt{MinSup} and \texttt{MinConf}. Since this process can extract a vast amount of rules, additional measures such as \textit{lift} are applied to filter out irrelevant rules~\citep{fournier2017survey, dm-concepts2011, handbook-edm2010}.

\subsubsection*{Sequential Pattern Mining}
\label{sec:background:seq}

\textit{Sequential pattern} is a frequently occurring subsequence in a given set of sequences~\citep{fournier2017survey, handbook-edm2010}. For example, it can be a progression of certain commands that many students used. Contrary to ARM, SPM can analyze data in which the ordering of items is relevant.

Again, sequential patterns are mined based on a \texttt{MinSup} threshold. To find a manageable amount of patterns, it is recommended to use algorithms that mine \textit{closed sequential patterns}~\citep{fournier2017survey, fournier2014fast, fumarola2016clofast}.

\subsection{Clustering}
\label{sec:background:clustering}

\textit{Clustering} is the process of assigning data points into groups called \textit{clusters} based on their similarity. Data in one group are similar to each other and dissimilar to data from other groups~\citep{madhulatha2012overview}. For example, in our context, we can group students based on the similarities in their command-line usage. Clustering is an unsupervised machine learning technique, so it does not use previously labeled data to assess new data. Instead, it organizes unlabeled data into \enquote{bundles}.

We focus on \textit{density-based} clustering, which defines a cluster as an area with a high density of data points; low-density areas separate individual clusters. Unlike \textit{partitional} clustering methods, such as the popular $k$-means clustering~\citep{lloyd1982least}, density-based approaches are better at recognizing arbitrarily shaped clusters and filtering noise or outliers. However, not all data points may end up in a cluster~\citep{beyer1999nearest, aggarwal2001surprising}.

\section{Related Work}
\label{sec:related-work}

This section reviews the publications related to the analysis of educational data. It also explains how our research differs from state of the art.

\subsection{Pattern Mining in Educational Data}

Association rule mining (ARM) or sequential pattern mining (SPM) has been employed to investigate various aspects of education. These include learner difficulties, correlations between learning behaviors and performance, and teaching strategies that lead to better learning~\citep{romero2020, bienkowski2012}.

\cite{handbook-edm7} applied ARM on data capturing students' usage of a learning management system, discovering relationships between students' activities and final grades. Instructors can use this information to adjust the course or identify struggling students early. \cite{kobayashi2014} also used ARM to uncover the errors that frequently co-occurred at various proficiency levels when learning spoken English. The pattern mining revealed types of mistakes that distinguish lower-level and upper-level students.

\cite{malekian2020} applied SPM on data representing students' actions and task submissions in an online learning environment. The researchers wanted to discover the behavior patterns that lead to successful or unsuccessful assessment outcomes. Therefore, they split the sequences of actions into two categories depending on the outcome of the sequence's final submission. The failed sequences contained mainly repeated assessment submissions and discussion forum views. In contrast, the passed sequences included multiple reviews of lecture materials. This information can be used to modify the learning environment to discourage unproductive behavior.

\cite{gao2021} mined sequential patterns from programming logs to identify struggling students. Timely recognizing these students is essential for promoting their learning. To establish ground truth, the researchers again split the logs of high- and low-performing students. Then, they mined patterns that either dominated in one group to discover its specifics, or occurred in both groups to reveal similarities. After that, they used the patterns as features in a classifier algorithm to predict student performance.

\subsection{Clustering of Educational Data}

\cite{handbook-edm6} motivate the usage of clustering in educational contexts. In addition, they also provide a brief overview of literature where clustering was applied to solve educational problems. Next, \cite{romero2010educational} and \cite{dutt2017systematic} performed literature reviews of EDM papers. Clustering has been used to provide feedback to instructors, detect undesirable or unusual student behavior, analyze and model student behavior, and group students by various characteristics, such as their learning approaches.

\cite{yin2015clustering} used the OPTICS algorithm to cluster students' programming assignments, aiming to support autograding based on the type of solution. Student source code was represented as an abstract syntax tree, with the normalized tree edit distance as the similarity measure for clustering. The researchers discovered clusters corresponding to distinct types of solutions (canonical, correct but longer code, complex solution, and so on).

\cite{mcbroom2016mining} mined submission logs from an autograding system for program code. They clustered weekly submissions to find approaches to each assignment while also analyzing the long-term behavior to learn how students develop. The researchers detected common behavioral patterns as early as in week three of the semester, and students' behavior largely remained the same. Teachers can use the gained insight to intervene when a student belongs to the cluster with a higher risk of failure.

The goal of \cite{piech2012modeling} was to study how students learn to program. To do so, the researchers captured and clustered temporal traces of student interactions with a compiler. They applied a hidden Markov model to the temporal traces and visualized it as a state machine for the cluster. The model then predicted student performance.

\cite{emerson2020cluster} explored novices' misconceptions in block-based programming. The researchers used logs of unsuccessful student attempts at programming assignments. The students' programs were represented by three families of features: basic block features, counts of specific block sequences, and the number of interactions with the system. The results revealed three clusters of students: exploratory, disorganized, and near-miss.

In their follow-up work, \cite{Wiggins2021} analyzed novices' hint requests in block-based programming. When a student asked for a hint, the time elapsed from the assignment's start and the percentage of code completion were recorded. Clustering of this data revealed five different groups of students based on their hint-taking strategies. For example, those that asked for a hint early and had low code completeness probably needed a \enquote{push} to start. Instructors can use this information to target the students' needs specific to the given group.

\subsection{Using Data for Student Assessment in Cybersecurity}
\label{subsec:related-work-cybersecurity}

\cite{Maennel2020} performed a thorough literature review of data sources that can serve as evidence of learning in cybersecurity exercises. These data sources include timing information, command-line data, counts of events, and input logs. Our paper investigates the applicability of \textit{command-line data} in educational assessment. Such data are collected in multiple state-of-the-art learning environments for cybersecurity training~\citep{weiss2017, Andreolini2019, Labuschagne2017, Tian2018}.

Weiss et al. demonstrated that command-line data from cybersecurity training are valuable for student assessment. They incorporated information about the students' exact steps, rather than just a numerical score indicating success or failure. They analyzed the students' work processes and the utilized command-line tools. Based on the command histories, they generated progress models of student approaches~\citep{weiss2016, weiss2017, my-2022-SIGCSE-modeling} and predicted their success~\citep{vinlove2020}. 

\cite{mirkovic2020} collected and analyzed command-line input and output from participants in hands-on cybersecurity exercises. The analysis system automatically compared the collected data with pre-defined exercise milestones and produced statistics about the participants' progress. It helped identify difficult sections of the exercises and students needing assistance, providing useful information to instructors.

\cite{abbott2015} parsed a dataset of logs from cybersecurity training into meaningful blocks of activity and statistically analyzed them. \cite{mcclain2015} further explored this dataset combined with questionnaires measuring the participants' experience in cybersecurity. They discovered that more experienced participants used specialized and general-purpose tools, while the less experienced participants focused only on specialized cybersecurity tools.

Finally, several works investigated the assessment of teams in sophisticated cyber defense exercises. \cite{Granaasen2016} collected network and system logs to study the performance of teams. Similar data sources were used by \cite{Henshel2016predicting} to assess and predict team performance. \cite{Maennel2017} proposed a systematic approach: a methodology to employ exercise data for team assessment. In contrast, we focus on individual assessment during exercises in the scope of classroom teaching.

\subsection{Summary of the Related Work}
\label{subsec:rw-summary}

Pattern mining and clustering were applied in educational contexts with interesting results. They can reveal students' misconceptions, approaches to solving the tasks, and behavioral patterns. These insights can improve educational assessment and feedback and target instruction to support students' needs.

The novelty of our paper is exploring these methods in the context of cybersecurity training. Previously, command-line data from cybersecurity training were analyzed using other methods, such as statistics, regular expression matching, and classifiers. We seek to discover insights gathered from cybersecurity training data using pattern mining and clustering, as well as demonstrate their usefulness for assessment. Moreover, we aim to uncover in-depth insights, not only assess the correctness of the student solution.

\section{Research Methods}
\label{sec:methods}

\begin{figure}[!ht]
\centering
\includegraphics[width=0.95\textwidth]{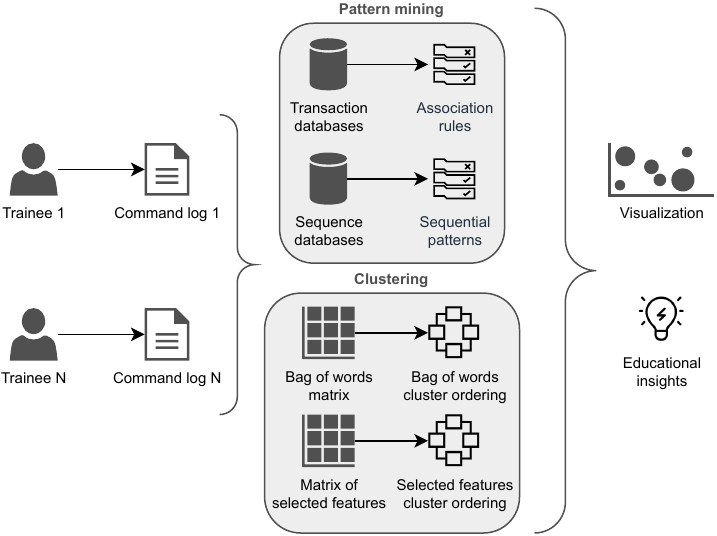}
\caption{The command logs collected from students act as input for pattern mining and clustering. The results are visualized and interpreted in \Cref{sec:results}.}
\label{figure:methods}
\end{figure}

This section explains the methods chosen to answer the research questions posed in \Cref{subsec:goal}. A visual overview of these methods is provided in \Cref{figure:methods}. In previous projects~\citep{TkacikThesis, PopovicThesis}, we prototyped the methods on smaller datasets, yielding initial results that we updated for this paper.

\subsection{Cybersecurity Training}

Our research analyzes data from cybersecurity training. Specifically, we focus on offensive security skills training in a sandboxed network emulated within an interactive learning environment. The following text introduces essential aspects of the training to provide context for the research.

\subsubsection*{Interactive Learning Environment}

The virtual machines for the training were hosted in \kypo~\citep{kypo-website, my-2021-FIE-kypo-csc}, which is a cloud-based infrastructure for emulating complex networks. For some training sessions, we alternatively used \creator~\citep{csc-website, my-2021-FIE-kypo-csc}: a tool for creating lightweight virtual labs hosted locally on the trainees' computers. This choice of the underlying infrastructure did not affect the training content, and the data collection was also equivalent.

Both platforms are open-source~\citep{my-2021-FIE-kypo-csc}, and cybersecurity instructors can freely deploy them for their purposes.

\subsubsection*{Training Format}

The trainees worked with the interactive learning environment either remotely via a web browser or locally on their computers. Each trainee accessed their own isolated sandbox containing a virtual machine with Kali Linux~\citep{kali-website}: an operating system distribution tailored for penetration testing that provided the necessary tools. The trainees completed a sequence of assignments presented via a web interface. Almost all the assignments were solved using command-line tools, which are described below.

The participants were allowed to use any sources on the Internet. Moreover, the interactive learning environment offered optional hints, which the trainees could reveal to get help with the current task. The usage of hints and outside help was allowed since the trainees were not evaluated summatively (that is, the training was not a graded exam). Instead, we focused on formative assessment and helping the students explore new cybersecurity skills.

\subsubsection*{Training Content}

Each trainee participated in exactly one of two types of training. Both trainings involved attacking an intentionally vulnerable virtual host using well-known security tools, but the trainings slightly differed in their content. In Training A (72 participants), the following tools were crucial: \texttt{nmap} for network scanning, Metasploit for exploitation, \texttt{john} for password cracking, and \texttt{ssh} for remote connection. Training B (41 participants) used \texttt{nmap} and \texttt{ssh} as well, but not Metasploit or \texttt{john}. Instead, it featured \texttt{fcrackzip} for cracking passwords to ZIP files (see \Cref{fig:1b-bruteforce,fig:1c-help}). None of the trainees was previously familiar with any of these two trainings.

Again, the training content is publicly available~\citep{our-games}. Training A corresponds to the cybersecurity game \textit{Secret laboratory} and its derivatives, while Training B corresponds to the game \textit{Junior hacker training}. Cybersecurity instructors can freely deploy these games in their classes and recreate the conditions for our research.

\subsubsection*{Training Participants} 

From August 2019 to February 2021, we hosted \numsessions\ cybersecurity training sessions for a total of \numtrainees\ trainees. Each training session usually took two hours to complete, and most of them were held remotely due to COVID-19 restrictions. The participants included:
\begin{itemize}
    \item undergraduate and graduate students of computer science from various European universities,
    \item high school students attending the national cybersecurity competition, and
    \item cybersecurity professionals.
\end{itemize}
They all attended voluntarily because of their interest in cybersecurity and were not incentivized. Although the participants do not form a random sample, we argue that it is practically infeasible to recruit a randomized population for this type of research. Therefore, we instead worked with the representatives of the target group for this cybersecurity training.

\subsubsection*{Ethical and Privacy-Preserving Measures for Research}

Since we carried out research with human participants, we ensured that the trainees would not be harmed in any way. We minimized the extent of data collection to gather only the data necessary for the research. We also received a waiver from our institutional ethical board since we do not collect any personally identifiable information.

The participants provided informed consent to the collection and usage of their data for research purposes. The collected data were thoroughly anonymized not to reveal the trainee's identity. As a result, it is impossible to track the trainee throughout future training sessions.

\subsection{Data Collection}

While the trainees solve the assignments, our infrastructure~\citep{my-2021-FIE-logging} automatically collects their submitted commands and the associated metadata. We gathered data from command-line tools in the Linux Bash terminal and Metasploit shell, which is software for penetration testing~\citep{msf-website}. These data, which are published (along with other training data) in an open-source article~\citep{Svabensky2021dataset}, serve as the input for pattern mining and clustering. We did not collect data from tools with a graphical user interface.

\subsubsection*{Data Format}

The command history of each trainee is captured in a single JSON file. The file consists of dozens of log records (78 per trainee on average), such that each record represents a single command executed by the trainee. \Cref{fig:command-json} shows an example of such a log record.

\begin{figure}[!ht]
\begin{center}
\small
\begin{verbatim}
    {
        "timestamp"  : "2020-07-03T08:09:25+01:00",
        "username"   : "root",
        "hostname"   : "attacker",
        "ip"         : "10.1.135.83",
        "sandbox_id" : "1",
        "wd"         : "/home",
        "cmd"        : "nmap --help",
        "cmd_type"   : "bash-command"
    }
    
\end{verbatim}
\end{center}
\caption{A single log record from a command history of one trainee.}
\label{fig:command-json}
\end{figure}

Each log record has a fixed number of attributes. For our purposes, the most significant are:
\begin{itemize}
    \item \texttt{timestamp}, representing the time of the command's execution in the ISO 8601 format,
    \item \texttt{cmd}, which represents the full command (the tool and its arguments) submitted by the trainee, and
    \item \texttt{cmd\_type}, the application used to execute the command: either \textquote{bash-command} for the tools executed within Linux Bash terminal, or \textquote{msf-command} for Metasploit shell.
\end{itemize}

\subsubsection*{Data Properties}

We collected \numcommandhistories\ commands, which constitute the dataset for this research, over the period of 1.5 years. Although this sample is not massive in volume, it captures the trainees' interactions deeply and over prolonged periods. Therefore, it fulfills the prerequisites of the chosen data mining methods.

Hands-on cybersecurity training is usually held in a group of lower tens of participants. Therefore, we consider the \numcommandhistories\ commands to be sufficient for evaluating the two data mining methods. On average, this dataset corresponds to 78 commands per trainee within the 1--2-hour time frame, which is appropriate for the chosen training format.

For this research paper, we focus on data processing after the training ends. Nevertheless, the used methods are applicable during the training for real-time assessment as well.

\subsection{Pattern Mining}

To enable mining patterns from the command-line data, our analysis scripts written in Python automatically transformed the input data into the \textit{transaction} and \textit{sequence} databases described below. These databases are an internal representation of the input data, and they serve as the input for ARM and SPM algorithms, respectively. A key advantage of pattern mining is that the data preparation is the same for assessing any task from the training.

\subsubsection*{Transaction Databases}

We parsed the dataset of commands to create two transaction databases used as input for ARM. The \textit{command transaction database} represents each submitted command as a separate transaction, and its goal is to reveal different properties of command usage. Each transaction contains four items that represent the attributes of the command:
\begin{itemize}
    \item \texttt{tool}, the name of the submitted command (e.g., \texttt{nmap} or \texttt{ssh}),
    \item \texttt{args}, the command-line arguments supplied to the tool,
    \item \texttt{app}, either Bash shell (Linux terminal) or Metasploit,
    \item \texttt{gap}, the time difference between the current and the following command.
\end{itemize}

For example, the command from \Cref{fig:command-json} can become a single transaction \{\verb!tool = nmap!, \verb!args = --help!, \verb!app = bash!, \verb!gap = low!\}. To achieve better interpretability, the \texttt{gap} attribute was automatically discretized~\citep[p. 102]{handbook-edm2010}: divided into categorical classes from the set \{low, medium, high, undefined\}, since the exact value in seconds is not too important. We followed the method previously published by~\cite{mccall2019}. First, the \texttt{gap} value in seconds was computed for each command. Then, gaps exceeding the arbitrary maximum of 20 minutes were discretized to \textquote{undefined}. This resolved the cases of long periods of trainee inactivity. The interval cut-off points for \textquote{low}, \textquote{medium}, and \textquote{high} categories were computed based on the mean gap from all gaps not exceeding the maximum.

The second database, called the \textit{tool transaction database}, contains transactions with only two attributes: \texttt{tool} and \texttt{gap}. We merged the consecutive uses of the same tool (regardless of the arguments) into a single transaction. The \texttt{gap} represents the time difference between the first use of a tool and the next use of a different tool; the values were discretized as before. The motivation for creating this database was to determine the difficulty of using different tools. If a tool is associated with long gaps, it may indicate that the trainees were unfamiliar with this tool and had difficulties using it.

\subsubsection*{Sequence Databases}

Three sequence databases were created as input for SPM. All three had \numtrainees\ sequences (corresponding to the number of trainees and the command log files), differing only in the contained items.

The first database, called \textit{command sequence database}, consists of sequences of executed commands. Each item represents a single command, both the tool and its arguments. For example, a sequence from this database can look like this: \verb!nmap --help, nmap 1.2.3.4, nmap -p 1000 1.2.3.4!.

The second database, \textit{tool sequence database}, contains sequences of tools only. Data from both Bash and Metasploit applications are included in the first two databases. This allows discovering longer patterns, which more accurately reflect the trainees' progress.

The third database, \textit{application sequence database}, stores sequences of applications utilized by the trainees to execute commands. Its goal is to reveal a high-level overview of alternating between applications. This database contains only two unique items: \texttt{terminal}, which includes all the commands executed in the Bash shell, and \texttt{metasploit}. \Cref{tab:PM-databases-summary} shows the number of transactions/sequences and unique items in each of our databases.

\begin{table}[!ht]
\centering
\small
\caption{The number of transactions or sequences and unique items contained in each database (DB) for pattern mining, separated for both Training A and B.}
\label{tab:PM-databases-summary}
\begin{tabular}{cp{8.8mm}rcrp{5mm}rcr}
 \textbf{Database} & \multicolumn{4}{c}{\textbf{Transactions or seqs}} & \multicolumn{4}{c}{\textbf{Unique items}}\\
  & \multicolumn{4}{c}{\textbf{(Training A / B)}} & \multicolumn{4}{c}{\textbf{(Training A / B)}}\\
 \hline
 \textit{Command transaction DB}  & & 5700 & / & 3134 & & 1932 & / & 1092 \\
 \textit{Tool transaction DB}     & & 4167 & / & 2062 & &  369 & / &  155 \\
 \hline
 \textit{Command sequence DB}     & &   72 & / &   41 & & 2076 & / & 1155 \\
 \textit{Tool sequence DB}        & &   72 & / &   41 & &  365 & / &  151 \\
 \textit{Application sequence DB} & &   72 & / &   41 & &    2 & / &    1
\end{tabular}
\end{table}

\subsubsection*{Association Rule Mining}

For ARM, we used Apyori~\citep{apyori-github}, the Python implementation of the Apriori algorithm. The \texttt{MinSup} threshold was manually tuned for each database since there is no simple method to determine it. The threshold was initially set to higher values and then gradually lowered to 0.01--0.04 until we reached a sufficient number of patterns manageable for interpretation. This approach is suggested by~\cite{fournier2013minsup} since finding suitable values depends on the data and specific use case.

The \texttt{MinConf} threshold is generally easier to set, because the database's properties influence \texttt{MinSup} more heavily than \texttt{MinConf}~\citep{fournier2012topk}. Since we were interested in rules with higher confidence, we used higher \texttt{MinConf} thresholds of 0.5. In contrast, \texttt{MinSup} needed to be much lower to extract a sufficient amount of rules. This was probably because our transaction databases contained many unique items relative to the total amount of transactions. If there were fewer unique items, \texttt{MinSup} could have been increased.

\subsubsection*{Sequential Pattern Mining}

For SPM, we used an open-source data mining library SPMF~\citep{fournier2016spmf}. It provides optimized and documented implementations of more than 190 data mining algorithms~\citep{spmf-website} often used as benchmarks in research papers~\citep{fournier2016spmf}. We selected CloFast~\citep{fumarola2016clofast}, an efficient algorithm for mining closed sequential patterns. The \texttt{MinSup} threshold was experimentally set from 0.3 to 0.7.

\subsection{Clustering}

A popular density-based algorithm is \textit{OPTICS} (Ordering Points To Identify the Clustering Structure)~\citep{ankerst1999optics}, an improved extension of a widely-used DBSCAN algorithm~\citep{tang2016}. For a data point to belong in a cluster, it must have at least \texttt{MinPts} points within its radius.

The result of OPTICS clustering is a \textit{reachability plot}. On the x-axis, it sorts all data points in the order of processing based on their similarity. Values on the y-axis represent the distance of a point from a previous one. Several similar points form a valley representing a cluster, while spikes represent noise or outliers~\citep{ankerst1999optics}.

In our research, we first represented each command as a Python object with the following attributes: \textit{tool}, \textit{arguments}, \textit{application type}, and \textit{timestamp}, simplifying the record in \Cref{fig:command-json}. Then, we used the commands in two different feature matrices that later act as an input for clustering.

\subsubsection*{Bag of Words Matrix}

\textit{Bag of words} model is a standard technique for obtaining features from text~\citep{pelanek2018}. Each text document is represented by a set of words it contains and their count. In our case, the \enquote{document} is a command history, and each tool is a \enquote{word}. We disregarded the command's arguments since we would obtain too many unique features and impair the performance of the clustering algorithm.

\subsubsection*{Matrix of Selected Features}

While the \textit{bag of words} model captures the used commands, it does not consider other information available in the logs. Therefore, we selected five custom features to capture other insights into how the trainees progressed:

\begin{itemize}
\item \texttt{bash-count}, the number of submitted Bash commands. A small number may suggest that a trainee did not progress far in training. The high number may indicate using a trial and error approach. 
\item \texttt{msf-count}, the number of Metasploit commands a trainee used. Metasploit may be new for some trainees, and the high number of executed commands may indicate difficulties with this part of the training.
\item \texttt{avg-gap}, the average delay between two commands. Large gaps between commands may suggest the trainee did not understand how to use a tool and possibly looked for the information online. Small delays may indicate brute-force guessing.
\item \texttt{opt-changes}, the number of times trainee used the same tool twice in a row but changed the options or arguments. A high count may show the trainee's unfamiliarity with the tool or inability to use it.
\item \texttt{help-count}, the number of times trainee displayed help information or manual page for any tool. It may also indicate the trainees' unfamiliarity with the tool.
\end{itemize}

All features were standardized, namely scaled by their maximum absolute value~\citep{scikit-scale}. We also checked the Pearson correlation between features, as a high value may make them redundant. While there was a correlation of 0.85 between \texttt{bash-count} and \texttt{opt-changes}, we preserved both because they capture different properties. All other features were correlated less (the absolute values ranged from 0.20 to 0.66).

\subsubsection*{Clustering Analysis}

We chose the OPTICS algorithm to cluster our data. For calculating the distance between data points, we selected cosine similarity. This measure performs well on high-dimensional data and is often used to compute text similarity~\citep{shirkhorshidi2015comparison}. For example, the command \verb!nmap -sn -PS22 10.1.26.9! has the similarity of 0.6 with the command \verb!nmap --script=vuln 10.1.26.9! and approx. 0.32 with the command \verb!nmap --help!.

During the setup, OPTICS takes only one parameter \texttt{MinPts}: the minimum number of points required for cluster formation. Theory suggests setting the number to \textit{ln(n)}, where \textit{n} is the number of points in the dataset~\citep{birant2007st}. For our dataset, the recommended value should be close to $ln(113) \approx 5$, which we selected.

\section{Results and Discussion}
\label{sec:results}

This section answers the two research questions (RQ) about insights gathered from pattern mining and clustering. We visualize and interpret the findings from specific training sessions and subsequently compare the two approaches.

\subsection{RQ1: Pattern Mining}
\label{subsec:results-rq1}

We now describe and discuss the results revealed by ARM and SPM.

\subsubsection*{Transaction Databases}

The \textit{command transaction database} revealed 51 association rules for Training~A and 50 for Training~B. \Cref{tab:arm_table1} presents the selected rules marked as interesting by measures such as lift. The first row shows that in Training~A, 64\% of commands executed in Metasploit had small gaps (delay times). This can mean that using Metasploit involved a rapid sequence of simple commands, or that the trainees experimented with a trial and error approach. The high support of the rule (23\%) can also indicate the overuse of Metasploit because it was needed only for one task in this training.

\begin{table}[!ht]
\begin{center}
\caption{Association rules mined from \textit{command transaction database} for Training A (rules A$x$) and Training B (rules B$x$). The antecedent and consequent of each rule are separated by \textquote{$\to$}. \textit{Sup} and \textit{Conf} stand for support and confidence rounded to two decimal places.}
\label{tab:arm_table1}
\begin{tabular}{clcc}
 \textbf{\#} & \textbf{Rule} & \textbf{Sup} & \textbf{Conf} \\
 \hline
 A1 & \texttt{APP=metasploit} $\to$ \texttt{GAP=low} & 0.23 & 0.64 \\
 A2 & \texttt{ARGS=[]} $\to$ \texttt{APP=terminal} & 0.20 & 0.66 \\
 A3 & \texttt{ARGS=[]} $\to$ \texttt{GAP=low} & 0.20 & 0.64 \\
 A4 & \texttt{ARGS=[] GAP=low} $\to$ \texttt{APP=terminal} & 0.15 & 0.74 \\
 A5 & \texttt{ARGS=[] APP=terminal} $\to$ \texttt{GAP=low} & 0.15 & 0.71 \\
 A6 & \texttt{GAP=medium} $\to$ \texttt{APP=terminal} & 0.12 & 0.64 \\
 A7 & \texttt{GAP=high} $\to$ \texttt{APP=terminal} & 0.11 & 0.69 \\
 B1 & \texttt{APP=terminal} $\to$ \texttt{GAP=low} & 0.64 & 0.64 \\
 B2 & \texttt{ARGS=[]} $\to$ \texttt{GAP=low} & 0.21 & 0.69
\end{tabular}
\end{center}
\end{table}

Generally, tools without arguments were associated with small gaps and often with Bash terminal commands. This most likely implies that tools without arguments are easier and faster to use. On the other hand, if a tool had medium or large gaps, it was used in the Bash terminal as well. This is because Bash offers many tools with various difficulty levels, some of which offer a multitude of options.

The \textit{tool transaction database} provides further insight into the tool usage. Tools such as \texttt{cd}, \texttt{ls}, and \texttt{cat}, as well as Metasploit commands (\texttt{use}, \texttt{set}, \texttt{show}) were associated with small gaps. However, \texttt{nmap} was associated with large gaps in 72\% of cases. This can indicate its difficulty of use or the long duration of the scan, which depends on the used arguments, as previously observed by~\cite{weiss2016}.

\subsubsection*{Sequence Databases}

The \textit{command sequence database} in Training~A revealed that trainees performed the Metasploit exploitation in various ways. Some steps were optional or performed in arbitrary order. When multiple approaches to a solution are possible, instructors can use this insight to show different examples in class, assess all the correct sequences as passed, or even discover novel solutions. Alternatively, when unsuitable subsequences are found, the trainees can be notified, corrected, or even penalized.

In Training~B, SPM showed that most trainees established an SSH connection only on the second or third try. When students learn error-prone actions, instructors should leave room for trial and error and not penalize the students for repeated tries. On the other hand, about a third of the trainees excessively used the \texttt{ls} tool (as much as 17 times within a single sequence, interleaved by other tools). Instructors should discourage unproductive behavior and maybe offer hints to students when such sequences are observed.

\begin{landscape}
\begin{figure}[!ht]
\centering
\includegraphics[width=\paperwidth]{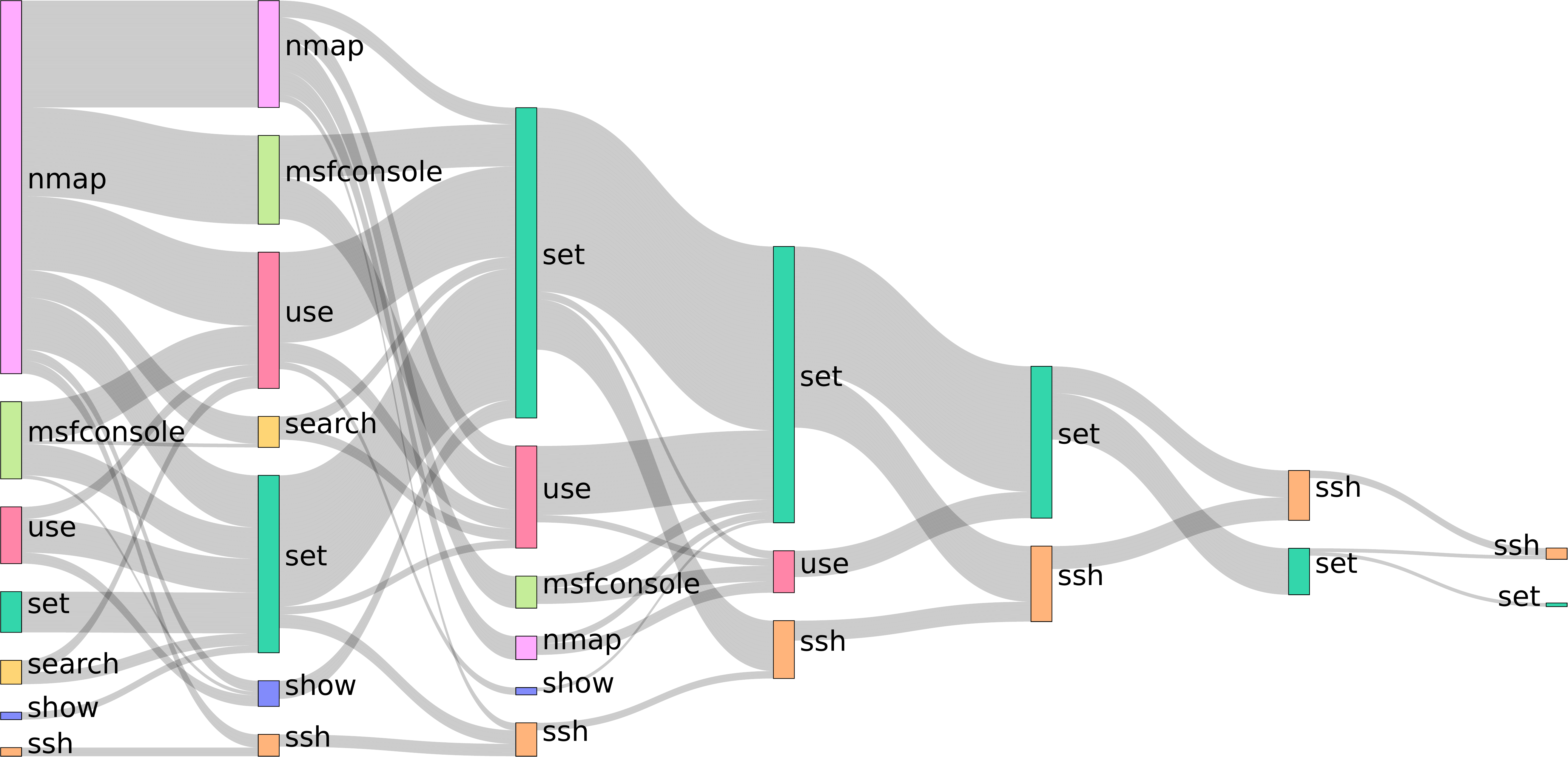}
\caption{A Sankey diagram of closed sequential patterns mined from the \textit{tool sequence database}. The commands \texttt{search}, \texttt{use}, \texttt{show}, and \texttt{set} were used in Metasploit; the others in Bash.}
\label{figure:sankey_sp_tools}
\end{figure}
\end{landscape}

The patterns from the \textit{tool sequence database} show that in Training~A, the participants usually progressed as instructors expected. They started with an \texttt{nmap} scan and proceeded with the Metasploit exploitation. This is visualized in \Cref{figure:sankey_sp_tools} using a \textit{Sankey diagram}. Nodes represent the items of the discovered patterns. Edges between the nodes represent subsequences of the patterns. The thicker the edge, the higher the support of the pattern in which the subsequence occurs.

The canonical solution featured these steps in the following order:
\begin{itemize}
    \item \verb!nmap <target>! -- scan the target IP address to discover available services;
    \item \verb!search <keyword>! -- find Metasploit exploits suitable for the discovered service based on the provided keyword;
    \item \verb!use <exploit>! -- select the correct exploit;
    \item \verb!show options! -- display parameters of the exploit that need to be set;
    \item \verb!set <option>! -- configure the exploit parameters (used three times to set three mandatory options);
    \item \verb!run! or \verb!exploit! -- execute the exploit script.
\end{itemize}

\Cref{figure:sankey_sp_tools} shows that most trainees did not use \texttt{search}, which suggests they received a hint about which exploit to use. This hint was available in the training platform. Moreover, few of them used \texttt{show} to display the exploit options. Instead, they started configuring them immediately, which again suggests they received a hint about which parameters the exploit has and how to configure them. Since the training offered an option to take hints, these actions were legitimate in our context.

In Training~B, the longest patterns feature sequences of \texttt{ls} and \texttt{cd} tools. This can indicate that the trainees struggled to find the files necessary to advance in the task. Again, instructors can provide hints to help the students who become stuck.

Finally, the \textit{application sequence database} confirms our intuition that in Training~A, the trainees did not often alternate between Bash and Metasploit. Instead, they used them in longer sequences. For example, the longest discovered pattern features 7 Bash commands, then 8 Metasploit commands, and then 5 Bash commands. The support of this pattern is 0.71, meaning that 71\% of trainees behaved this way.

In Training~B, the sequence of 12 Bash commands has the support of 0.98, meaning that all but one trainee executed at least 12 commands. If a trainee uses too few commands, it may indicate issues with the assignment, a surprisingly effective solution, outside help, or even cheating.

\subsubsection*{Limitations of Pattern Mining}

Setting the \texttt{MinSup} and \texttt{MinConf} parameters must often be done by trial and error, since there is no universal guide. Also, pattern mining algorithms extracted relatively many patterns, many of which were trivial, for example, \texttt{TOOL=cd} $\to$ \texttt{APP=terminal}.

Defining the importance of patterns can address this problem. For example, a pattern describing relationships between \texttt{tools} would be more important than the usage of the terminal (\texttt{app}) itself. Alternatively, additional postprocessing can remove trivial patterns to save the analyst's time. A text-based file defining uninteresting patterns can be used to filter the patterns.

Finally, it can be difficult to interpret why certain patterns occur. Additional information and context are needed to maximize the usefulness of extracted patterns.

\subsubsection*{Summary of RQ1}

ARM and SPM are suitable for uncovering the following educational insights:
\begin{itemize}
    \item Approaches to solving the tasks, namely typical associations of tools and their arguments (for ARM) or sequences of commands (for SPM).
    \item Mistakes and errors based on incorrectly used tools or unknown commands and sequences.
    \item Problematic tasks within the training, such as when a student attempts to use a tool several times in a row (for SPM).
    \item Novel solutions, such as when unexpected but correct tools appear in a rule or a sequence.
    \item Tools used at the beginning or toward the end of the task, based on whether the sequences often begin or end with a certain tool (for SPM).
    \item Frequency of tools' usage, such as the commands utilized by most trainees or overuse of a tool, which is proportional to the rule's or sequence's support.
    \item Timing information, namely small or large gaps between two submissions of commands associated with certain tools (for ARM).
\end{itemize}
The results of pattern mining can be tabulated (see \Cref{tab:arm_table1}) or visualized in a Sankey diagram, such as the one in \Cref{figure:sankey_sp_tools}.

\subsection{RQ2: Clustering}
\label{subsec:results-rq2}

Now, we continue with the results of clustering the \textit{bag of words} and \textit{selected features} matrices.

\subsubsection*{Bag of Words Cluster Ordering}

When clustering the 72 trainees from Training A, 31~trainees form Cluster 1, 14 trainees constitute Cluster 2, 5 trainees form Cluster 3, and 22 were designated as outliers. Cluster 2 is the most compact because of low reachability distance between the points. This implies that the trainees progressed strongly similarly.

\begin{figure}[!ht]
\centering
\includegraphics[width=\textwidth]{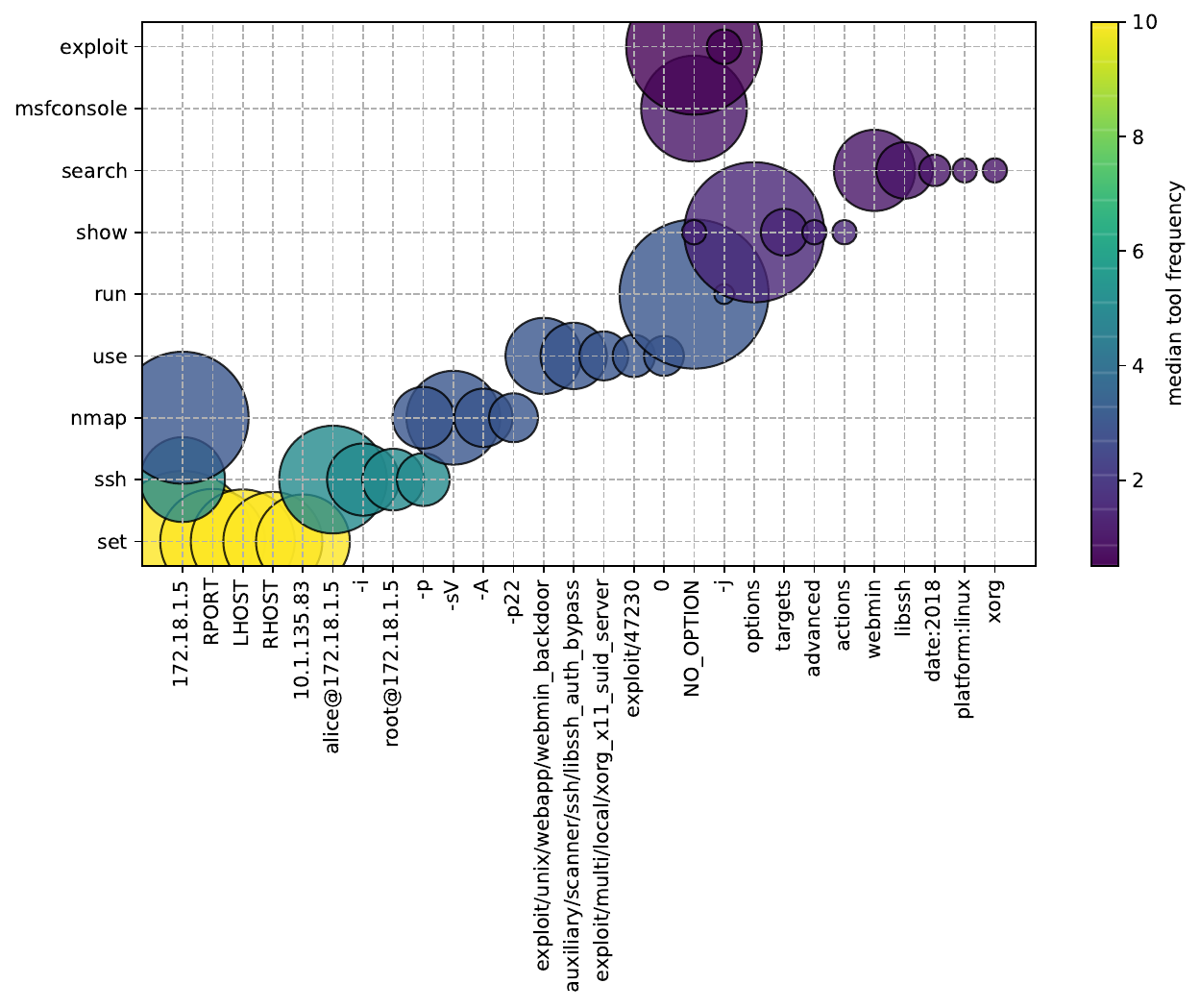}
\caption{Bubble plot showing the most common \texttt{tool} and \texttt{argument} combination for Cluster~1. The size of the bubble is correlated with the argument frequency. The color represents the median tool frequency for the cluster.}
\label{figure:cluster1_commands}
\end{figure}

When visualizing the most common combinations of tools and arguments (see \Cref{figure:cluster1_commands}), we discovered that Cluster 1 trainees used \texttt{nmap} and \texttt{ssh} with certain arguments slightly more often. For example, they executed a correct \texttt{nmap} scan multiple times, maybe to assure themselves of the results.

Trainees from Clusters 1 and 3 also experimented much more with setting the Metasploit exploit options, and attempted to search for and use several different exploits. Cluster 2 trainees selected and configured the suitable exploit on fewer tries. A relatively low number of Metasploit commands and the lack of option variety suggest that Cluster 2 trainees did not struggle with Metasploit. Instructors can use this information to check in with Cluster 1 and 3 trainees and ask them whether they are stuck or need assistance.

Cluster 2 trainees used more Bash commands on average, and they used commands for changing and listing directories (\texttt{cd} and \texttt{ls}) overwhelmingly more often. They probably had trouble locating the files crucial for the task. Based on this insight, instructors can again provide targeted help to trainees in this cluster.

For the 41 trainees in Training B, the clustering was not too fruitful. 13 trainees formed a cluster, while the remaining 28 were designated as outliers. The trainees in a cluster again used a lot of \texttt{cd} and \texttt{ls} tools, and experimented with \texttt{scp} and \texttt{fcrackzip} more often than the remaining trainees.

Examining the trainees designated as outliers can also yield interesting results. One of the outliers did not use \texttt{nmap} for network scanning, but \texttt{ike-scan} and then \texttt{zenmap}. This shows that alternative tools are possible for solving the tasks, and outliers can still be successful in the training. It is worth noting that even small differences or deviations in a single task can be enough for the trainee to be considered an outlier. However, these cases have to be further investigated manually.

Finally, another outlier brute-forced the searching of argument combinations for \texttt{john}. Therefore, outliers can also be trainees behaving problematically.

\subsubsection*{Selected Features Cluster Ordering}

Clustering of Training A data formed four clusters: with 6, 9, 14, and 6 members, respectively. Cluster 1 had the largest count of Metasploit commands and the smallest average time gap between their submitted commands. These differences were also confirmed by pairwise t-tests statistically significant at $p\leq 0.01$. Since only a few Metasploit commands were needed to reach the solution, this indicates a trial-and-error approach of trainees in this cluster. Moreover, these trainees did not display the manual pages or the tools' usage help. Such unproductive behavior can be automatically recognized, along with notifying the instructors. On the other hand, Cluster 4 trainees displayed command help the most often, which can be suitable while learning.

Cluster 2 and Cluster 3 behaved in an almost opposite ways. The former used the most Bash commands with relatively small gaps, and the latter used the least Bash commands with the largest gaps. This suggests that Cluster~3 trainees did not progress far, perhaps due to lack of motivation or skill.

Two clusters emerged in Training B. The first had fewer submitted commands with larger gaps. The second submitted many commands with small gaps and changed the command arguments often. These trainees probably struggled to figure out the correct argument combination.

Overlaps with the \textit{bag of words} clusters were minimal, suggesting that the results largely depend on the chosen features. Both approaches can provide useful insights; however, as in almost all machine learning approaches, it is difficult to select the best features.

\subsubsection*{Limitations of Clustering}

The main limitation of clustering is that determining relevant features is hard. In addition, the relatively small sample size was problematic for the chosen clustering algorithm. Sometimes, only one cluster was formed, or only a few data points belonged to a cluster. Nevertheless, the format of the training implies that massive amounts of command histories cannot be collected.

\subsubsection*{Summary of RQ2}

Clustering can reveal the following educational insights:
\begin{itemize}
    \item Similarities and differences between trainees' approaches to the training, for example, in typically used combinations of tools.
    \item Alternative solutions to training tasks based on examining outliers.
    \item Behavioral patterns, such as help-seeking or submitting many commands in a rapid succession.
\end{itemize}
The results of clustering and the associated features can be easily visualized or tabulated, which provides a straightforward overview (see, e.g., \Cref{figure:cluster1_commands}).

\subsection{Comparison of the Two Approaches}
\label{subsec:results-cmp}

We used two approaches, \textit{pattern mining} and \textit{clustering}, to analyze data from cybersecurity assignments completed via a command line. \Cref{tab:observations} summarizes the different situations that can occur during the training and are of interest to instructors. Then, \Cref{tab:comparison} provides a grand overview of insights discoverable with the two approaches. Not all of them were demonstrated by our data, but they can be investigated by future research.

\begin{table}[!ht]
\centering
\footnotesize
\caption{The insights and situations that can arise during the training.}
\label{tab:observations}
\begin{tabular}{ll}
    \textbf{Positive or neutral observations} & \textbf{Negative observations} \\
    \hline
    P1 The trainee is proficient & N1 The trainee lacks skill \\
    P2 The trainee is motivated & N2 The trainee is demotivated \\
    P3 The trainee progresses smoothly & N3 The trainee experiences difficulties \\
    P4 The trainee received allowed help & N4 The trainee received prohibited help \\
    P5 The trainee corrected a mistake & N5 The trainee uses a trial-and-error approach \\
    P6 The trainee discovered a novel solution & \\
    P7 The trainee is taking a break & \\
    P8 The tool executes quickly & N8 The tool executes slowly \\
    P9 The tool is easy to use & N9 The tool is difficult to use \\
    & N10 The tool is used too little \\
    & N11 The tool is used too much \\
    P12 The task features simple commands & \\
    & N13 The task is not designed clearly 
\end{tabular}
\end{table}

\begin{table}[!ht]
\centering
\footnotesize
\caption{The comparison of results of pattern mining (ARM, SPM) and clustering (C). The column \textit{Explanations} refers to possible causes in \Cref{tab:observations}.}
\label{tab:comparison}
\begin{tabular}{llll}
    \textbf{Method} & \textbf{Insight category} & \textbf{Result} & \textbf{Explanations} \\
    \hline
    ARM, C & Solution approaches & Typical combinations of tools & task-dependent \\
    SPM & Solution approaches & Typical sequences of tools & task-dependent \\
    all & Tool use frequency & Low rule support / use count & P6, N1, N10 \\
    & & High rule support / use count & N1, N5, N11, N13 \\
    ARM, C & Timing information & Low command delay & P1, P4, P8, P9, P12, N4 \\
    & & High command delay & P7, N1, N2, N3, N8 \\
    C & Trainee similarities & Common behavioral patterns & task-dependent
\end{tabular}
\end{table}

These insights can also occur in combination. For example, a low frequency of command usage combined with large gaps probably suggests demotivation or lack of skill.

\subsubsection*{Similarities of Pattern Mining and Clustering}

Both pattern mining and clustering can reveal trainees' strategies utilized to solve the tasks. These include desirable solutions, mistakes and errors, and novel approaches. They can also highlight statistical properties of the solutions, such as frequently used tools or their time gaps.

Methodically, the process of pattern mining and clustering is relatively straightforward. As long as the input data format and constraints are preserved, it is sufficient to use existing implementations of these algorithms.

Insights from both pattern mining and clustering can be targeted at specific trainees. Instructors can provide suitable feedback to the whole cluster of students or all students whose logs matched a specific pattern. This way, instructors can help the struggling trainees if they exhibit signs typical for low-performing clusters or associated with undesirable patterns.

\subsubsection*{Differences of Pattern Mining and Clustering}

Setting the initial parameters appears to be easier for the OPTICS clustering algorithm compared to ARM and SPM. OPTICS recommends setting \texttt{MinPts} approximately to the natural logarithm of the sample size. For pattern mining, the \texttt{MinSup} and \texttt{MinConf} parameters need to be set experimentally. Nevertheless, clustering requires a careful selection of features, which can include the collected data as well as properties derived from them.

Density-based clustering is more prone to small sample size. On the contrary, our dataset of thousands of commands was sufficient for pattern mining. In fact, most public datasets previously used for ARM contain thousands up to a million transactions, and datasets for SPM start with as little as ten sequences up to ten thousand~\citep{fournier2021datasets}. (However, these datasets come from other domains, such as word corpora or clickstream data from websites.) As a result, educational researchers who have just begun to collect data may experience the cold start problem. Especially when using clustering, their early dataset will not be large enough to provide insights about the first few students.

Finally, the results of clustering are more easily interpretable. Pattern mining can yield a large number of trivial patterns. Nevertheless, interpreting the clustering results requires further investigation of the properties of the discovered clusters. Patterns are readable directly.

\subsection{Comparison With Related Work}

\Cref{sec:related-work} reviewed numerous approaches to student assessment and usage of pattern mining and clustering to analyze educational data. We now compare our methods and results with those presented in related work in order to highlight our contributions.

One novel aspect of our research is the application domain of cybersecurity, since most of the related work focused on other areas, such as programming education~\citep{gao2021, yin2015clustering, mcbroom2016mining, piech2012modeling, emerson2020cluster, Wiggins2021}. Educational researchers and practitioners in cybersecurity and related domains, such as operating systems and networking, may benefit from the presented evaluation featuring authentic cybersecurity training data.

Several learning environments for cybersecurity allow logging command-line interactions~\citep{my-2021-FIE-logging, mirkovic2020, Andreolini2019, Labuschagne2017, Tian2018}, although this practice is still relatively rare. Nevertheless, even if interesting datasets are acquired, few methods have been explored for their automated analysis (see \Cref{subsec:rw-summary}).

Student assessment in cybersecurity was specifically reviewed in \Cref{subsec:related-work-cybersecurity}. Based on inspecting the related work, we believe this is the first study evaluating the applicability of pattern mining and clustering algorithms on cybersecurity training data. Other works focused, for example, on generating progress models of students~\citep{my-2022-SIGCSE-modeling}, predicting their success~\citep{vinlove2020}, or assessing team performance~\citep{Granaasen2016, Henshel2016predicting, Maennel2017}.

\subsection{Educational Implications}
\label{subsec:results-implications}

Our research demonstrated the automation of discovering educational insights. Previously, these insights had to be revealed manually by the instructor, which was time-consuming, or were even completely unavailable. In particular, the insights gained from pattern mining and clustering have the following implications for teachers, educational researchers, and other stakeholders:
\begin{itemize}
    \item \textit{Classroom-wide instruction} -- instructors can show the typical or rare solution approaches to the students and discuss them together. They can also explain the erroneous or novel solutions. If students are aware of examples of good or bad practices, they can follow or avoid them, respectively.
    \medskip
    
    For example, in our data, \texttt{nmap} was often associated with high time gaps. Instructors can revisit the explanation of this tool and stress how its argument combinations affect the scan duration.
    \medskip
    
    \item \textit{Targeted instruction} -- when a student exhibits patterns associated with errors or unproductive behavior, the instructor can intervene appropriately. This intervention can include providing tailored hints, feedback, scaffolding, or suitably correcting the student. Identifying struggling students early and helping them is crucial for supporting their learning.
    \medskip
    
    For example, we discovered a cluster of trainees who adopted a trial-and-error approach when using Metasploit. If this happened during class, instructors could visit these students in real-time and provide suitable assistance. By identifying specific students belonging to the cluster, the instructor can save time by providing the same help to all students in that cluster.
    \medskip
    
    \item \textit{Marking/grading} -- knowing the common mistakes aids with both manual and automated grading. Instructors can create a grading rubric based on the observed errors and approaches. Moreover, an autograder can be set up to grade specific actions as passed or failed.
    \medskip
    
    For example, the trainees who used a correct sequence of Metasploit commands to configure all exploit steps can be awarded a point.
    \medskip
    
    \item \textit{Task design} -- based on summarizing the common approaches of students and discovering novel approaches to the solution, instructors can design more suitable tasks. This includes fixing unclear or problematic tasks.
    \medskip
    
    For example, we discovered that in Training B, participants excessively used the \texttt{cd} and \texttt{ls} tools to traverse the filesystem. The assignment can specify more clearly what type of file to look for and where. It can also feature hints.
    \medskip
    
    \item \textit{Machine learning} -- the patterns and their features (such as using a particular command or making a specific mistake) can act as input in other machine learning models for further analysis and student modeling.
    \medskip
    
    For example, the discovered patterns could be used to train a classifier to predict student success or failure.
    \medskip
    
    \item \textit{Curricular support} -- several policies and curricular guidelines in cybersecurity~\citep{cc2020, csec2017, Parrish2018} prescribe \textit{what} skills should be taught and assessed. However, the information about \textit{how} to perform this assessment is left to the educators. Our paper demonstrates a possible solution for assessing hands-on exercises in cybersecurity, which other educators can adopt or adapt.
\end{itemize}

\section{Conclusion}
\label{sec:conclusion}

Automated student assessment in cybersecurity is becoming more and more relevant. Finding better ways to analyze data from cybersecurity training is needed to support more effective hands-on training. Yet, this research area is still in its early stages.

To contribute to the body of knowledge on student assessment, we investigated automated methods for analyzing log data from authentic educational contexts. We mined \numcommandhistories\ commands from several-hours-long training sessions with small groups of computing students. Then, we discussed the observations relevant for instructors and researchers in cybersecurity and beyond.

Our results include the prototype implementation, evaluation, and comparison of two data mining approaches within specific cybersecurity training, as well as general insights and lessons learned. Answering our two research questions revealed that:
\begin{enumerate}
    \item \textit{Pattern mining} is suitable for revealing solution approaches of students, their misconceptions, and difficult training tasks.
    \item \textit{Clustering} highlights similarities and differences between approaches of students, grouping them based on their behavioral patterns.
\end{enumerate}

Other educators can use these insights to improve cybersecurity training in their context or adapt them to training in other domains. Pattern mining and clustering are suitable for any problem-solving assignments that yield interaction data. Instructors can exploit these data to identify and redesign problematic sections of the training, reveal new solutions to the tasks, and provide targeted instruction and feedback to trainees.

\subsection{Practical Contributions and Supplementary Materials}

In addition to educational implications described in \Cref{subsec:results-implications}, we share numerous artifacts with the community of instructors, researchers, and developers. These artifacts enable replicating our study setup and advancing the research in cybersecurity education. Moreover, the tools are applicable for hands-on security classes. These artifacts are open-source and include:
\begin{itemize}
    \item Cybersecurity training content~\citep{our-games}, which can be deployed in either of our learning environments: \kypo~\citep{kypo-website} and \creator~\citep{csc-website}. Cybersecurity instructors can freely use them to host cybersecurity training sessions~\citep{my-2021-FIE-kypo-csc}.
    \item Logging infrastructure~\citep{my-2021-FIE-logging}, which enables researchers to collect command-line data like for this paper.
    \item The analyzed dataset, which has been published with records from other trainings~\citep{Svabensky2021dataset}. Since each record is a command submitted by a person, accumulating these data is a challenge on its own. Therefore, such datasets are rare and may help other researchers.
    \item The created software that applies pattern mining and clustering on the data. It includes a Python implementation of extracting and visualizing the patterns and clusters. This implementation can serve as a starting point for the developers of learning environments when integrating the researched methods~\citep{zenodo}.
    \item Visualizations and the full results~\citep{zenodo}.
\end{itemize}

\subsection{Future Work}

This research offers many possibilities for extension. In pattern mining, transaction databases can include time-related information, such as the duration of running a command. This would distinguish a difficult task from a command that took a long time to execute. Sequential databases can include timestamps to describe time gaps between sequences in a pattern. Additionally, the dataset can be expanded with information about other actions of trainees, such as asking for a hint. As a result, we would discover sequences that preceded help-seeking. Finally, we can consider the whole command history of a student as a single transaction to generate new types of insights.

Enhancements are possible for clustering as well. One is the clustering of time series: each training would be represented as time series of commands encoded as vectors. The other option is to use different algorithms, such as hierarchical clustering.

Yet another extension is to incorporate live data mining during an ongoing training. Online algorithms can provide relevant insights to instructors in real-time. Their results could also be used as a basis for a recommender system that would provide hints for stuck trainees. If a trainee needs help, the system could recommend a hint that helped another trainee from the same cluster.

\backmatter

\section*{Statements and Declarations}

\bmhead{Funding}
This research was supported by the ERDF project \textit{CyberSecurity, CyberCrime and Critical Information Infrastructures Center of Excellence} (No. CZ.02.1.01/0.0/0.0/16\_019/0000822).

\bmhead{Acknowledgments}
We thank Radek Pelánek and Tomáš Effenberger for their valuable feedback that helped with the early stages of this paper. We also thank the anonymous reviewers who provided useful perspective to improve the final version of the paper.

\bmhead{Competing interests} The authors have no competing interests to declare.

\bmhead{Availability of data and materials} The accompanying software code and results are published in a Zenodo repository~\citep{zenodo}. The dataset was published in a separate data article~\citep{Svabensky2021dataset} (this paper used a subset of the published data).

\bmhead{Authors' contributions}
\textit{Valdemar Švábenský}: Conceptualization, Methodology, Validation, Investigation, Resources, Data Curation, Writing -- Original Draft, Visualization, Supervision, Project administration. \textit{Jan Vykopal}: Investigation, Resources, Writing -- Review \& Editing. \textit{Pavel Čeleda}: Writing -- Review \& Editing, Funding acquisition. \textit{Kristián Tkáčik}: Software, Formal analysis, Data Curation, Writing -- Review \& Editing, Visualization. \textit{Daniel Popovič}: Software, Formal analysis, Writing -- Review \& Editing, Visualization.

\bibliography{references}

\end{document}